\documentclass[11pt,twocolumn,twoside,a4paper,amsmath,amssymb,aps,showkeys,showpacs]{revtex4}

\usepackage{amsfonts}
\usepackage{graphics} 
\usepackage{epsfig}
\usepackage{fancyheadings}

\newcommand{\pt}{p_\mathrm{T}}
\newcommand{\kt}{k_\mathrm{T}}
\newcommand{\et}{E_\mathrm{T}}
\newcommand{\gsim}{\,{\buildrel > \over {_\sim}}\,}
\newcommand{\gev}{\mathrm{GeV}}
\newcommand{\gevc}{\mathrm{GeV}/c}
\newcommand{\Raa}{R_\mathrm{AA}^{jet}}
\newcommand{\RAA}{R_\mathrm{AA}}

\begin{document}
\thispagestyle{myheadings}
\rhead[]{}
\lhead[]{}
\chead[Jan Kapit\'an]{Jets in heavy ion collisions at RHIC}

\title{Jets in heavy ion collisions at RHIC}

\author{Jan Kapit\'an}
\email{kapitan@rcf.rhic.bnl.gov}


\affiliation{%
Nuclear Physics Institute, ASCR, Na Truhlarce 39/64,
18086 Praha 8, CZECH REPUBLIC }%





\received{ ? }

\begin{abstract}

Full jet reconstruction in heavy-ion collisions enables a complete study of the modification of jet structure due to energy loss in hot and dense QCD matter, but is challenging due to the high multiplicity environment.
The STAR and PHENIX collaborations at RHIC have recently presented measurements of fully reconstructed jets from p+p, Cu+Cu and Au+Au collisions at $\sqrt{s_\mathrm{NN}} = 200~\mathrm{GeV}$.
We review the first results on inclusive jet spectra, di-jets and fragmentation functions and discuss their implications on understanding of jet quenching.

\end{abstract}

\pacs{12.38.Mh , 21.65.Qr , 25.75.Bh}

\keywords{ jet finding, recombination algorithms, background subtraction, 
jet broadening, fragmentation function }

\maketitle

\renewcommand{\thefootnote}{\fnsymbol{footnote}}
\renewcommand{\thefootnote}{\roman{footnote}}


\section{Introduction}
\label{intro}

Jets are remnants of hard-scattered partons, which are the fundamental objects of perturbative QCD. At Relativistic Heavy Ion Collider (RHIC), they can be used as a probe of the hot and dense matter created in heavy ion collisions. Interaction and energy loss in the medium lead to jet quenching in heavy ion collisions. Until recently, jet quenching was studied indirectly using single particle spectra and di-hadron correlations~\cite{quenching}. These measurements are however limited in the sensitivity to probe partonic energy loss mechanisms due to biases toward hard fragmentation and small energy loss~\cite{trenk}.

Developments in theory (for example~\cite{fj,bgsub}) and experiment (detector upgrades, increased RHIC luminosity) finally enabled full jet reconstruction in heavy ion collisions~\cite{initial}. Full jet reconstruction reduces the biases of indirect measurements and enables access to qualitatively new observables such as energy flow and fragmentation functions. 

As a baseline measurement for heavy ion jet studies, p+p collisions at the same energy are used. To isolate initial state effects from medium modification, the STAR collaboration investigated nuclear $\kt$ broadening~\cite{vitev} in d+Au at $\sqrt{s_{NN}} = 200~\gev$ collisions~\cite{epshep}. This study indicates that initial state effects for jet production are rather small.

After description of new jet finding techniques suitable for heavy ion environment, we review the recent results on jet spectra, di-jets and fragmentation functions from $\sqrt{s_{NN}} = 200~\gev$ Cu+Cu and Au+Au collisions by the PHENIX~\cite{YShiQM09,YShiDPF} and STAR~\cite{MP,EB} collaborations.

\section{Jet reconstruction algorithms}
\label{jets}
The aim of jet reconstruction is to recover initial hard-scattered parton energy by measuring the products of its fragmentation. Both the STAR and PHENIX detectors are equipped with tracking detectors, which are used to measure charged component of jets and electromagnetic calorimeters that measure neutral component of jets (mainly $\pi^{0}$). A few particle species (namely (anti-)neutrons and $K^{0}_{L}$) are undetected and this effect has to be included in jet energy scale correction. Further experimental details can be found in~\cite{YShiQM09,YShiDPF,MP,EB}.

A jet algorithm that is infrared and collinear (IRC) safe is needed for meaningful jet definition~\cite{snowmass}. One example are the recombination jet algorithms kt and anti-kt, parts of the FastJet package~\cite{fj}, which are used by the STAR collaboration. The PHENIX collaboration uses a jet finder based on Gaussian filter~\cite{gauss_original,gauss}, that is also IRC safe. It also naturally suppresses background by giving more weight to the core of a jet and is well suited for a limited acceptance detector such as PHENIX central arms. 

\section{Heavy ion background}
\label{background}

Once jets are defined, the large underlying event background in heavy ion collision has to be taken into account. For example, in central Au+Au collision at RHIC, the average background under a jet with $R=0.4$ is $\approx 45~\gev$ ($R$ is cone radius for a cone algorithm, or resolution parameter for recombination algorithm).

It was shown by STAR, that in p+p collisions at $\sqrt{s} = 200~\gev$ $\approx 85(95) \%$ of jet energy for jet $30 < \pt < 40~\gevc$ is contained within a cone of $R = 0.4 (0.7)$~\cite{helenQM09}. The background under a given jet behaves like $R^{2}$, therefore relatively small (such as $R = 0.4$) cone radii / resolution parameters are used for jet finding in heavy-ion collisions.
Another way of reducing background, which comes mainly from soft particle production, is to introduce a low $\pt$ cut on particles used for jet finding. However, this method introduces fragmentation biases and is therefore generally avoided.

Large and not completely uniform background gives rise to many false jets, structures unrelated to hard scattering, identified by jet finder as jets. These can be either quantified and subtracted statistically, or identified on jet-by-jet basis~\cite{fakejets}. 

Background increases apparent jet energy and due to background fluctuations it is not possible to determine the background under a given jet exactly. The average background can be subtracted on jet-by-jet basis given a jet area and an average background density ($\pt$ per unit area) in given event. Background fluctuations smear jet energy, and this smearing has to be unfolded statistically to be able to compare to p+p collisions. 

The STAR collaboration subtracts false jets statistically, using either events randomized in azimuth with jet-leading particles removed (for inclusive jet spectrum analysis), or jet spectrum at 90 degrees in azimuth with respect to the leading jet in the event (for di-jet and fragmentation function analysis).    
To subtract the background energy under given jet, a method based on active jet areas~\cite{bgsub} is applied jet-by-jet: $\pt^{Rec} = \pt^{Candidate} - \rho \cdot A$, with $\rho$ estimating the background density in given event and $A$ being the jet active area. 
Background fluctuations are assumed to be Gaussian with width obtained by embedding p+p jets simulated by Pythia~\cite{pythia} into Au+Au events. Unfolding is performed to unsmear these background fluctuations.

A fake rejection discriminant~\cite{gauss} based on Gaussian-weighted sum of $\pt^{2}$ is used by the PHENIX collaboration to reject false jets on jet-by-jet basis.
Unfolding based on embedding p+p jets into Cu+Cu events is used to correct jet energy for background effects~\cite{YShiQM09}.

\section{Inclusive jet spectra}
\label{sec_spectra}
Nuclear modification factor of jets $\Raa$ is defined as the ratio of jet spectra in A+A collisions and in p+p collisions scaled by $\langle N_{bin} \rangle$. 
$\Raa$ is expected to be unity for unbiased jet reconstruction with possible small deviations due to initial state effects (such as EMC effect~\cite{emc}).

\begin{figure}[htb]
\centering
\includegraphics[width=8.2cm]{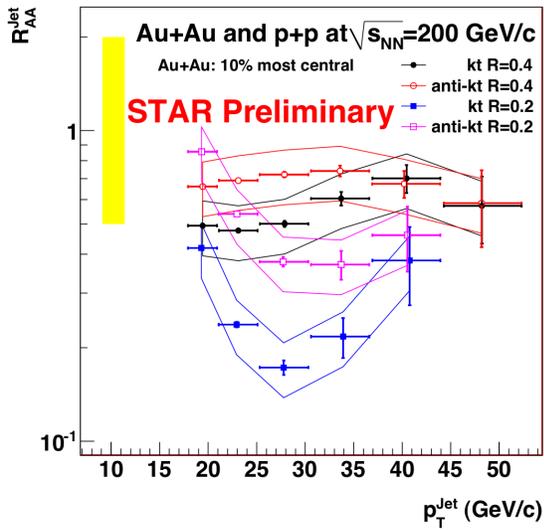}
\caption{$\Raa$ in Au+Au collisions by STAR, yellow band shows jet energy scale uncertainty. Taken from~\cite{MP}.}
\label{fig:Raa}
\end{figure}

\begin{figure}[htb]
\centering
\includegraphics[width=8.2cm]{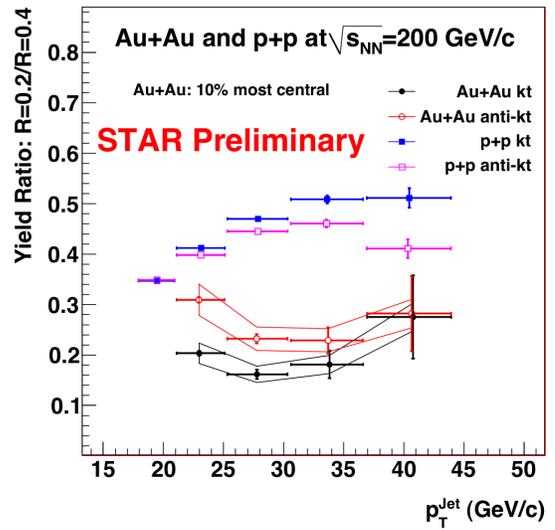}
\caption{Ratio of jet yields for $R = 0.2, 0.4$ in p+p and Au+Au collisions by STAR. Taken from~\cite{MP}.} 
\label{fig:Rratio}
\end{figure}

The STAR collaboration measured $\Raa$ in 0-10\% most central Au+Au collisions~\cite{MP}. The results for two values of resolution parameter ($R = 0.2$, $R = 0.4$) and two jet finding algorithms (kt, anti-kt) are shown in Figure~\ref{fig:Raa}. 
For $R = 0.4$, $\Raa$ is compatible with unity with large uncertainties, while for $R = 0.2$, the jets are significantly suppressed. The differences between kt and anti-kt algorithms are most likely due to different sensitivity to heavy ion background. 
Figure~\ref{fig:Rratio} shows the ratio of jet $\pt$ spectra for $R = 0.2$ over that for $R = 0.4$ separately for p+p and Au+Au. The ratio is strongly suppressed in Au+Au with respect to p+p, indicating a substantial broadening of the jet structure in heavy ion collisions~\cite{MP}.

\begin{figure}[htb]
\centering
\includegraphics[width=8.2cm]{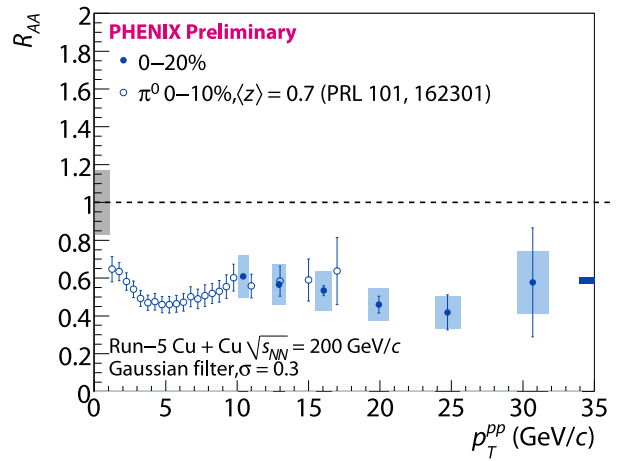}
\caption{$\Raa$ in 0-20\% most central Cu+Cu collisions (solid symbols) and $\pi^{0}$ $\RAA$ in 0-10\% most central Cu+Cu collisions (open symbols). By PHENIX, taken from~\cite{YShiDPF}.}
\label{fig:phenixRaa}
\end{figure}

The nuclear modification factor of jets in 0-20\% most central Cu+Cu collisions by the PHENIX collaboration~\cite{YShiDPF} is shown in Figure~\ref{fig:phenixRaa}. These results were obtained using a Gaussian filter with $\sigma = 0.3$. The jets show significant suppression, similar to that of pions.

\section{Di-jet and fragmentation function measurement}
Insight into jet structure is provided by studying the jet fragmentation function (FF). In-medium softening of the FF with respect to p+p reference should be observable in Au+Au for an unbiased jet population~\cite{softening}. 

The STAR collaboration presented measurement of fragmentation functions in p+p and 0-20\% most central Au+Au collisions~\cite{EB}. Recoil jets on the away side of an electromagnetic calorimeter triggered jet (a single tower with $\et > 5.4~\gev$) were used to maximize the medium path length and therefore possible FF modification. 

Anti-kt algorithm with a resolution parameter $R = 0.4$ was used. The jet $\pt$ is not corrected for instrumental effects yet and is marked $p_\mathrm{T, rec}$. The recoil jet energy was determined using a resolution parameter $R = 0.4$, whereas charged hadrons in a larger cone ($R = 0.7$) around the jet axis were used to construct the FF. Background contributing to the FF was subtracted event-wise using the charged hadron $\pt$ distribution in out of cone area.

\begin{figure}[htb]
\centering
\includegraphics[width=8.2cm]{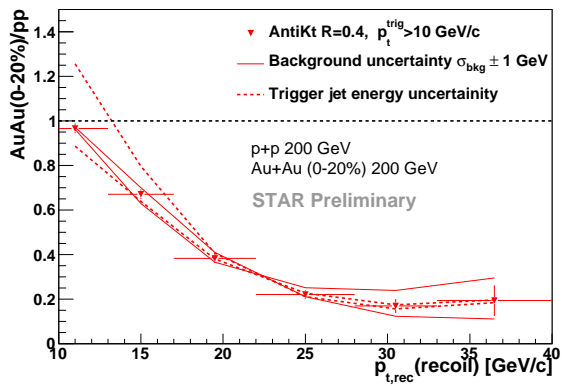}
\caption{Ratio of recoil jet $\pt$ spectra in Au+Au to p+p by STAR. Taken from~\cite{EB}.}
\label{fig:dijet}
\end{figure}

\begin{figure}[htb]
\centering
\includegraphics[width=8.2cm]{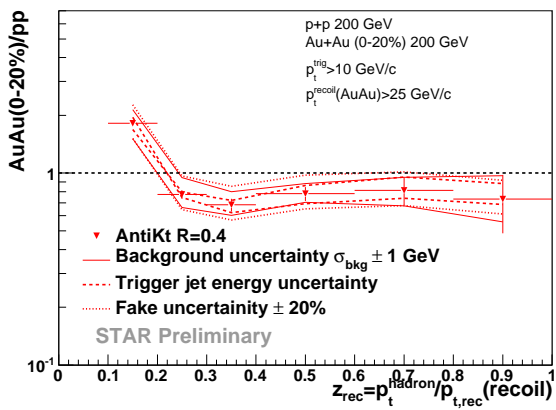}
\caption{Ratio of $z_\mathrm{rec} = p_\mathrm{T}^\mathrm{hadron} / p_\mathrm{T, rec}^\mathrm{recoil}$ distributions in Au+Au to p+p by STAR. Taken from~\cite{EB}.}
\label{fig:FF}
\end{figure}

Figure~\ref{fig:dijet} shows the ratio of Au+Au to p+p di-jet spectra, indicating a strong suppression of recoil jets in 0-20\% most central Au+Au collisions. This is in contrast to the expected value of unity for unbiased jet reconstruction. Figure~\ref{fig:FF} shows the ratio of Au+Au to p+p recoil jet FF for $p_\mathrm{T, rec}^\mathrm{recoil} > 25~\gevc$. No strong modification of the fragmentation function for $z_\mathrm{rec} \gsim 0.2$ is observed.

\section{Discussion}
\label{discussion}

Given different collision systems and centralities, the STAR and PHENIX results on $\Raa$ can not be quantitatively compared. However, they can be individually compared to pion $\RAA$. As can be seen in Figure~\ref{fig:phenixRaa}, PHENIX $\Raa$ in Cu+Cu collisions is consistent with $\RAA$ of pions.
In case of the STAR results on Au+Au collisions (Figure~\ref{fig:Raa}), $\Raa$ for $R = 0.4$ is significantly higher (less suppressed) than pion $\RAA$ ($\approx 0.2$ for $\pt > 5~\gevc$~\cite{singleRaa}).

PHENIX $\Raa$ is clearly below unity and the difference is likely beyond what is expected from EMC effect. This could be generally explained by either an energy shift toward lower energies (due to possible jet broadening), or the highly quenched jets may not be recovered by jet finder or not pass the false jet rejection algorithm. STAR $\Raa$ for $R = 0.4$ is compatible with unity, but a relatively large suppression is not ruled out. The difference between jet spectra ratios for $R = 0.2$ and $R = 0.4$ clearly indicates jet broadening in heavy ion collisions, which may lead to a shift of the reconstructed jet energy towards lower values, leading to decrease of $\Raa$.

The qualitative difference between STAR and PHENIX $\Raa$ may be due to different sensitivity of used jet algorithms to possible broadening of jet structure.
Also it may be due to different treatment of false jets between the two collaborations. The method currently used by STAR for jet $\pt$ spectra is close to a lower bound on false jet rate. On the other hand, PHENIX Gaussian filter approach may reject some heavily quenched jets although study using jet quenching model PYQUEN~\cite{pyquen} suggests the algorithm doesn't reject quenched jets~\cite{YShiDPF}.

STAR studies of dijets and fragmentation functions are consistent with the picture of jet broadening due to quenching. Significant suppression of recoil jets together with the absence of strong FF modification in Au+Au could be explained by broadening of the quenched recoil jets, causing an energy shift towards lower energies and artificial hardening of FF.
Alternatively, a part of recoil jet population may not be recovered due to strong broadening. In this case the measured FF in Au+Au may be that of surviving recoil jets, i.e. those with little modification (such as tangential emission or punch through jets)~\cite{EB}.

\section{Summary}
\label{summary}

By giving access to hard-scattered parton energy for the first time in heavy ion collisions, full jet reconstruction in $\sqrt{s_\mathrm{NN}} = 200~\gev$ Cu+Cu and Au+Au collisions brings qualitatively new insights into hot and dense QCD matter produced at RHIC.
Current results indicate broadening of jet structure due to quenching, but further studies of uncertainties due to jet finding algorithms and false jet rejection methods are needed to quantify the observed effects.

These new results will also be confronted to theory predictions (recent review of jet quenching models and related Monte Carlo codes can be found for example in~\cite{enterria}).
Theoretical and experimental progress achieved in full jet reconstruction in heavy ion collisions at RHIC is extremely important for heavy ion physics at CERN Large Hadron Collider (LHC), that is scheduled to start with p+p collisions at the end of 2009.

\section*{Acknowledgement}
\label{acknowledgement}

This work was supported in part by GACR grant 202/07/0079 
and by grants LC07048 and LA09013 of the Ministry 
of Education of the Czech Republic.


\label{last}
\end{document}